# KINETIC INVESTIGATION OF WATER-CATION INTERACTION IN CLINOPTILOLITE AT ROOM TEMPERATURE

*G. Carotenuto*

Institute for Polymers, Composites and Biomaterials (IMCB-CNR) - National Research Council. Piazzale E. Fermi, 1 – 80055 Portici (NA). Italy.

**Abstract.** The kinetics of water adsorption by a natural zeolite (clinoptilolite) sample has been investigated by high-frequency AC current intensity measurements. According to the achieved kinetic results, cations should play a relevant role in the clinoptilolite hydration, in fact most of adsorbed water stay in the cationic sites. The water molecules associate with cation one-by-one. In particular, the forced adsorption of water molecules in a wet atmosphere is a quite slow process, while water desorption in air or dry atmosphere is a spontaneous and fast process. The observed increase of cation mobility could be adequately explained by assuming an arrangement of water molecules between the cation and the negatively charged oxygens contained in the cationic site. Such molecular arrangement could increase the strength of both dipole-cation and hydrogen bond interactions.

## 1. Introduction

Zeolites are ionic conductors [1], that have the capability to adsorb small gaseous molecules that can permeate the regular array of channels present in their crystalline lattice [2]. Cations present in zolites (mainly alkali and earth alkaline metals) are involved both in the electrical conduction and in the adsorption mechanism. As a consequence, a variety of electrical chemo-sensors have been fabricated using zeolitic materials [3]. In addition, the zeolite adsorption properties have been exploited for other technological applications, like for example: fragrance encapsulation, therapeutic gas delivery, molecular traps and 'trapdoors', gas separation/purification and storage, desiccant/dryers, pollutant remover, etc. [4]. For all these applications the zeolite hydration level in service can be electrically evaluated, since the amount of cationic sites determines the adsorption capacity. Since zeolites contain also external silanols (Si-OH), molecular adsorption can take place also on these groups [5] in addition to cations. However, the silanol quantity in the zeolite framework is low because these groups are present mostly on the surface and rarely in the framework structure. In addition these groups are less effective than cations in adsorbing water and other polar/polarizable molecules, because the electric field of cations could act at very long distance. Since cations play a fundamental role in adsorption, it is very important to know the way polar molecules like water associate with cations and 3D-framework in the cationic site. Information on the organization of water in the cationic site can be achieved by kinetic investigation that can be performed by electrical measurements. In particular, the way cations influence the hydration/dehydration processes can be directly investigated by studying the temporal evolution of electrical transport in zeolite exposed to water vapor [6]. In fact, the adsorption of water molecules in the areas of cation sites influences the cation ability to move under the effect of an applied electric field, which reflects in the behavior of current intensity under a constant applied voltage. Consequently, an exact picture of mechanism involved in water adsorption process, that is the role of cations and the 3D-framework surrounding these cations in the adsorption, can be deduced by analyzing the electrical behavior of zeolite during water adsorption. Electrical measurements need biasing the sample with an adequate AC voltage of convenient frequency in order to avoid sample/electrode interface polarization phenomena. At high frequency (e.g., 5kHz), the current intensity variation can be directly correlated to variations in the carrier concentration, under constant applied voltage conditions. Therefore, the temporal behavior of the current intensity represents exactly the variation of the hydrated cations concentration. These electrical data can give kinetic information on the water adsorption and desorption processes, useful for the mechanism comprehension.

Here, the kinetics of the water adsorption/desorption reaction has been investigated by time-resolved measurements of the current intensity moving in the material biased with a constant AC sinusoidal voltage having a frequency of 5kHz. To selectively study the adsorption on cations, all non cationic sites (silanols) were first saturated by water using a hydration/dehydration cycle. This study was performed on a very common sample of natural zeolite (clinoptilolite), which has high natural abundance and is commercially available at low cost. In order to establish: phase composition, approximate Si/Al ratio, contained cation types, and water content, this material was also characterized by X-ray diffraction (XRD), scanning electron microscopy (SEM), energy dispersive X-ray spectrometry (EDS), and infrared spectroscopy (FT-IR).

## 2. Experimental

A commercial sample of natural clinoptilolite, provided by T.I.P. (Technische Industrie Produkte GmbH), was used in the as received form for time-resolved electrical measurements. Specimens for electrical characterization were fabricated by cutting the raw zeolite pieces in little slabs (5.0x5.0x3.0 mm) by a diamond saw.

Electrical measurements were performed with the two-contacts method. In particular, a curable silver paste (ENSON, EN-06B8), dried in air for 2 days and then backed in an oven at 140°C for 30min,was used to electrically contact the slab surface. To avoid sample/electrode interface polarization phenomena, the electrical transport in natural clinoptilolite samples was investigated by using high frequency alternate current measurements. A sinusoidal signal, produced by a DDS signal generator (GW Instek, AFG-2112), was applied to the sample and the resulting effective current intensity ($I_{eff}$) was measured by placing a 100kHz bandwidth True-RMS digital multimeter (Brymen, BM869s) in series with the sample. A sinusoidal signal of $20V_{pp}$ and 5kHz was used for the time-resolved measurements. While at low and medium frequencies the applied electric field causes both electric transport in the material and sample/electrode interface polarization, at high frequencies (e.g., above 1kHz) all produced hydrated cations are involved in the electrical transport and, consequently, the measured electrical current can be considered as directly proportional to the concentration of hydrated cations. This experimental condition is of a fundamental importance for the electrical monitoring of the cation hydration/dehydration reactions. In order to record current intensity during the time-resolved measurements, a devoted multimeter data logger system was used (software: Bs86x Data Logging System Ver. 6.0.0.3s). Measurements were done at room temperature and coaxial cables (BNC) were used for connections. To expose zeolite samples to a constant humidity environment, a plastic self-made cell was used. The saturated salt solution technique [9] was used for controlling humidity content in this cell. In particular, in order to achieve an atmosphere with 75% of humidity, which allowed easily measurable current intensity values (micro-amperes) an atmosphere with 75% of humidity was generated by using wet NaCl crystals placed at cell bottom (distilled water was used to wet NaCl). The cell contained a small hole to keep always the sample at 1 atm. A similar cell, containing freshly activated silica gel at bottom, was used to study the spontaneous dehydration process.

Si/Al ratio and type of contained ions were established by SEM (FEI Quanta 200 FEG microscope) and EDS analysis, respectively. The crystalline nature of the commercial clinoptilolite sample was investigated by XRD (X'Pert PRO, PANalytical). The clinoptilolite sample was also spectroscopically characterized by Furier-transform infrared spectroscopy (FT-IR), using a Nicolet apparatus in the ATR configuration. The thermal gravimetric analysis (TGA) was performed on the clinoptilolite sample by using a TA Instruments Q500, operating in nitrogen flow with a constant rate of 10°C/min.

## 3. Results and Discussion

The different crystalline phases present in the commercial sample of natural clinoptilolite were identified by X-ray diffraction (XRD). As visible in Figure 1, the XRD diffractogram contains the diffraction patterns of clinoptilolite, anortite, and stilbite. In particular, the most intensive characteristic peaks of clinoptilolite are visible at 9.8446°, 22.4026°, 30.0076°, 31.9576° [7,8]. The main peaks of anortite are visible at 21.9089, 27.7716°, 28.0837°, and stilbite has one peak placed at 19.0616°.

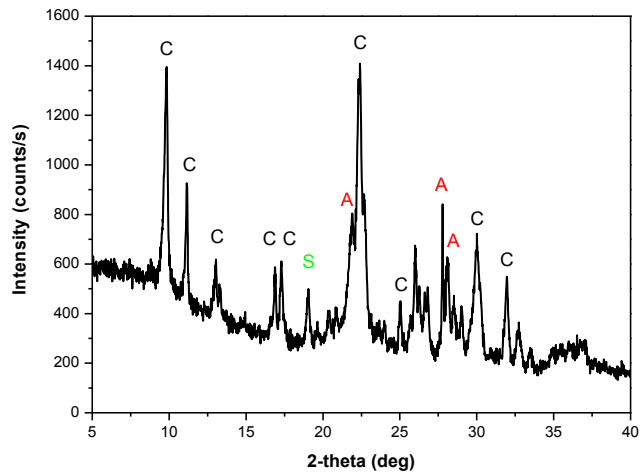

**Fig. 1 –** X-ray diffractogram of the studied natural clinoptilolite sample.

Extra-frame cations have been identified by EDS analysis (see Figure 2), and their approximate abundance has been estimated too [9]. According to the EDS analysis, the Si/Al ratio was 5.3.

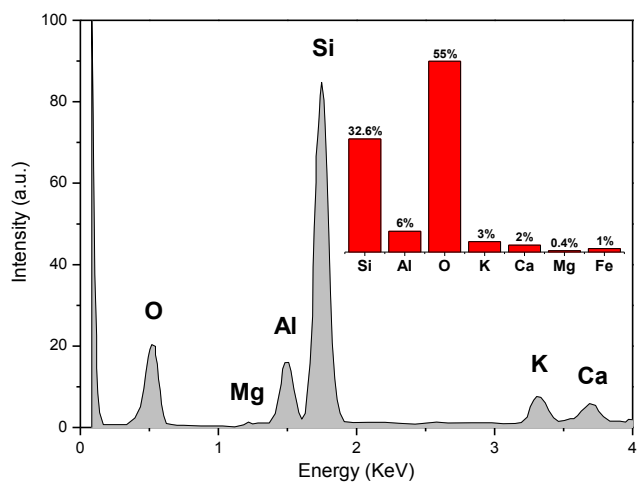

**Fig. 2 -** EDS spectrum of natural clinoptilolite sample.

The possibility for moisture to be adsorbed directly on cations (in addition to the hydroxyl groups, OH, present in zeolites as external silanols, Si-OH) has been proved by FT-IR investigations [10-14]. The FT-IR spectra of clinoptilolite in the O-H stretching and bending spectral ranges are shown in Figure 3. In particular, the analysis of the structure of the stretching absorption band reveals the presence of three components centered at 3610cm$^{-1}$, 3420cm$^{-1}$, and 3256cm$^{-1}$. The first absorption band is commonly attributed to the OH groups belonging to water molecules adsorbed on cations. This absorption band appears at a wavenumber lower than that of hydroxyls in gaseous water molecules (i.e., 3657cm$^{-1}$ (symmetric stretch) and 3756cm$^{-1}$ (asymmetric stretch), because of the strong polarization induced by the cation presence. In fact, when the water molecule is associated with a cation, the oxygen-hydrogen bond force constant (k) is lowered by inductive and resonance effects. Thus, the absorption band position changes according to the harmonic oscillator spectroscopic law: $\bar{n} = \frac{1}{2\pi c} \cdot \sqrt{\frac{k}{\mu}}$, where $\bar{n}$ is the wavenumber and k the Hooke's force constant of the O-H chemical bond and μ the corresponding reduced mass. However, this resonance is at a wavenumber higher than that of hydroxyl groups in liquid water (3400cm$^{-1}$), because for geometrical reasons hydrogen bonding results more effective than cation bonding in facilitate stretching vibrations by lowering the Hooke's force constant (k).

The adsorption bands at 3420cm$^{-1}$ and 3256cm$^{-1}$ are attributed to the OH groups belonging to silanols groups (Si-OH) located on the external surface of zeolite (terminal silanols). The first absorption band should be attributed to those silanols that are not interacting with water molecules. While, the absorption band observed as a shoulder in the FT-IR spectrum at 3256cm$^{-1}$ should belong to the OH groups of those silanols that interact with the adsorbed water molecules by hydrogen bonds.

Water molecules adsorbed in zeolite (both on cations and silanols) produce also a weak and sharp peak at 1628cm$^{-1}$, produced by the bending vibration [14]. The position of this signal is less sensible to the involvement of OH groups in physical interactions.

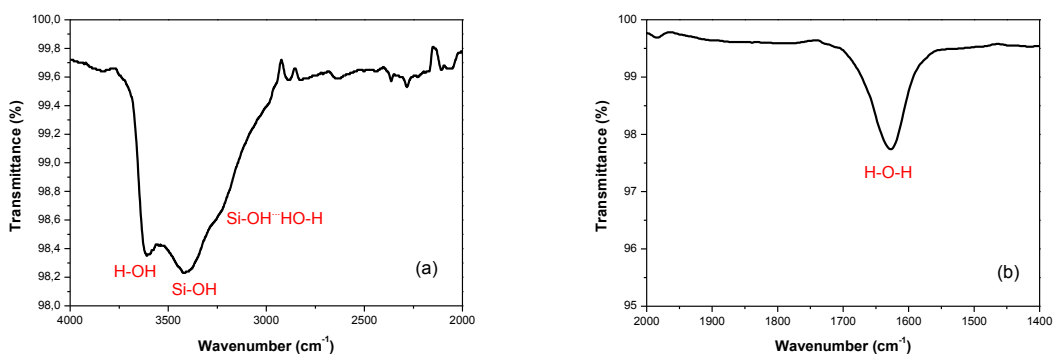

**Fig. 3 -** FT-IR spectra showing the stretching resonances of OH in silanols and in $H_2O$ associated with cations and silanols (a), and bending resonances of OH in H₂O associated and not associated with cations and silanols (b).

Zeolites has a strong ability to adsorb polar/polarizable molecules. These molecules can be adsorbed on two different types of sites: cations and hydroxyls (i.e., terminal silanols, Si-OH). The comprehension of the mechanism involved in the reversible adsorption of polar and polarizable molecules (like, for example, water) by cations in zeolites could be extremely useful for developing new advanced technological applications for these materials. In fact, the electrical conductivity of these materials is strongly influenced by the presence of molecules adsorbed by the cation sites and therefore they can be conveniently exploited as electrical chemosensors, etc. [15-17]. The adsorption of polar/polarizable molecules, like for example water, on the cation sites can be easily investigated by an electrical technique based on the measurement of the intensity of an AC current moving in the sample under a constant voltage [6]. However, since these molecules are adsorbed not only on cations, but also on the silanols, it is not easy to selectively investigate the mechanism of molecular adsorption on the cation sites by the electrical technique. In fact, when water adsorption is on both type of sites (i.e., absorption on dry zeolite or zeolite leaved in air for long time), a mass transport control has been found by the electrical investigation. In particular, the hydration of a dry sample (a new sample or a sample taken in air for a long time) showed a diffusive temporal behavior (parabolic, $I \propto \sqrt{t}$) of current intensity (see Fig. 4).

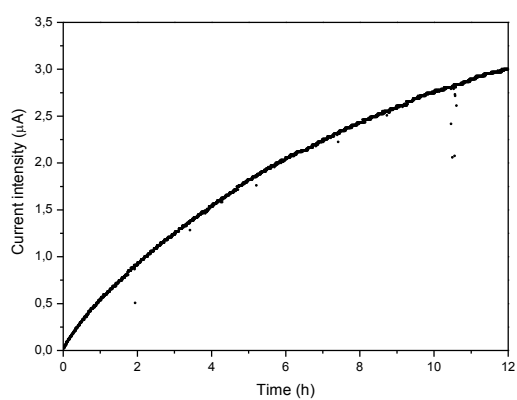

**Fig. 4 -** Temporal evolution of current intensity during the first sample hydration.

To achieve kinetic information on the absorption on cationic sites, a method based on electrical monitoring of absorption after saturation by water of all non-cationic sites has been developed (see Scheme I). Since water is promptly desorbed from cations while more time is required for desorption from the OH sites, it is possible to electrically study adsorption and desorption from the only cation sites. In fact, the electrical monitoring of re-adsorption just after a water adsorption/desorption cycle allows to investigate the adsorption on the only cationic sites. Instead, the electrical monitoring of the desorption process gives information on the only cationic sites because desorption from -OH takes much more time and do not affect the electrical behavior.

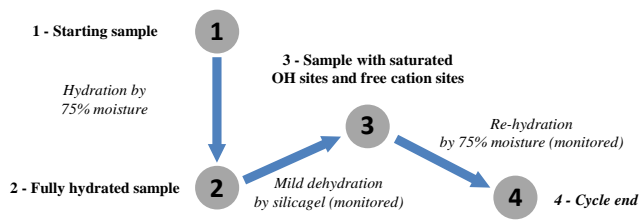

**Scheme I** – Dehydration/Re-hydration technique for the kinetic investigation of water adsorption on cations.

A time-resolved study of clinoptilolite-K electrical transport, based on biasing the sample with a high frequency (5kHz) sinusoidal signal, during the re-hydration and dehydration processes was used to investigate the mechanism involved in water adsorption/desorption by cations. For a generic $Me^+$ cation (in this case, $K^+$), the re-hydration reaction should be the following:

$$Me^+ + H_2O_{(g)} \rightleftarrows MeOH_2^+ \qquad (1)$$

Obviously, reaction (1) is reversible because the hydrated cations can spontaneously decompose giving back $Me^+$ and water. Two different freedom levels are possible for the extra-framework cations depending on their hydration state. Cations close to empty water sites have a high charge density (one or two charges in the ionic volume) and simultaneously interact with the partial negative charges on the aluminum and two oxygen atoms (see Scheme II). Such multiple and intensive electrostatic interactions strongly limits the movement of these cations under an alternate electric field. Differently, solvated cations are quite free to move and may undergo the effect of the AC electric field. In fact, the cation-site electrostatic interaction is lowered since their distance is increased. Finally, the hydration process deeply modifies the electrical behavior of clinoptilolite.

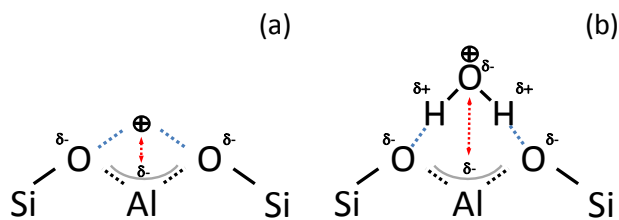

Scheme II–Structure of the cationic site in dry (a) and hydrated clinoptilolite (b).

Since only solvated cations can move under the effect of an electric field, the relative increase of current intensity ($\Delta I/I_0$) during the hydration process can provide useful kinetic information on the solvated cation formation reaction, provided that the voltage applied to the sample remains a constant during this process (to avoid changes in the cation speed). In fact, from the law of direct proportionality between current density and carrier (hydrated cations) concentration ($J = A \cdot [MeOH_2^+]$, with $A = q \cdot e \cdot \mu \cdot E$, where q is the cation charge number, e is the elementary charge: $1.6 \times 10^{-19}$C, E is the local electric field, and μ is a constant characteristic of the cation, known as mobility [18]) it follows that the temporal evolution of the relative current density increase is coincident with relative increase of the hydrated cations concentration:

$$\frac{\Delta I}{I_0} = \frac{\Delta J}{J_0} = \frac{\Delta[MeOH_2^+]}{[MeOH_2^+]_0} \qquad (2)$$

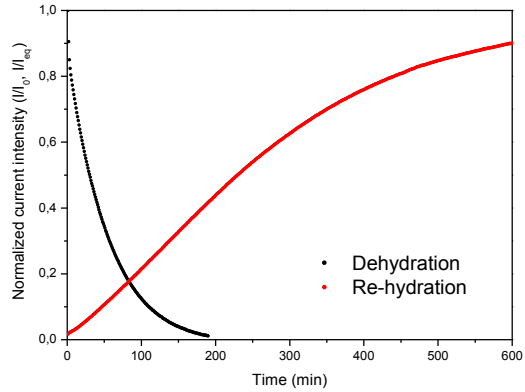

**Fig. 5 -** Temporal evolution of the normalized current intensity during dehydration and re-hydration.

The current intensity in clinoptilolite-K and therefore the carrier concentration was promptly affected by exposition to humidity, thus confirming the role of cations in the water adsorption mechanism. In addition, an exponential increase of the hydrated cation concentration was found (see Figure 5) by measuring the current intensity relative variation during re-hydration of clinoptilolite-K exposed to 75% moisture (sinusoidal signal with a frequency of 5kHz and amplitude of $20V_{pp}$). This temporal evolution of current intensity during the re-hydration process is compatible with the hypothesized reversible reaction in Eq. 1. In fact, according with the molecularity of this elemental reaction, the kinetic expression is the following:

$$\frac{d[MeOH_2^+]}{dt} = k[H_2O][Me^+]_0 - (k[H_2O] + k') \cdot [MeOH_2^+] \qquad (3)$$

where k and k' are the kinetic constants of the direct and inverse reactions at room temperature. After separation of variables, Eq. (3) can be integrated to give:

$$\ln\left(\frac{b+1-\frac{[MeOH_2^+]}{[MeOH_2^+]_0}}{b}\right) = -a \cdot t \qquad (4)$$

with: $a = k \cdot [H_2O] + k'$ and $b = \frac{k \cdot [H_2O] \cdot [Me^+]_0}{a \cdot [MeOH_2^+]_0} - 1$. It follows:

$$\frac{\Delta[MeOH_2^+]}{[MeOH_2^+]_0} = b \cdot (1 - e^{-a \cdot t}) \qquad (5)$$

At reaction beginning, Eq. (5) can be well approximate by a linear function (pseudo-zero kinetic order):

$$\frac{\Delta[MeOH_2^+]}{[MeOH_2^+]_0} \sim a \cdot b \cdot t \qquad (6)$$

Eq. (5) describes exactly the experimentally observed temporal behavior of the relative current intensity increase. Thus, the use of a moderately humid environment (i.e., 75% by weight) seems to avoid multiple adsorption of water molecules at the same cationic site.

Owing to the high humidity conditions used for the re-hydration experiment, the inverse reaction in Eq. (1) can be approximately neglected. In this case, the kinetic expression is similar to Eq. (5) but the a and b constants have the following values: a=k·[H$_2$O], and b=([Me+]$_0$/[MeOH$_2^+$]$_0$)-1. The kinetic constant for this simplified hydration process can be approximately calculated. In particular, the kinetic constant is obtained as the ratio between the $\Delta I/I_0$ curve slope at t=0 and the product between the curve asymptotic value and water concentration, expressed as weight fraction (0.75), and it results $k = (30.4 \pm 0.9) \cdot 10^{-4}$min$^{-1}$ at room temperature.

The spontaneous desorption of water molecules from clinoptilolite-K in dry air was monitored by recording the temporal evolution of current intensity inside the material under the same bias conditions (sinusoidal signal with frequency of 5kHz and amplitude of 20V$_{pp}$). In a dry environment, water molecules leave the cation surface, according with the following dehydration reaction, corresponding to the inverse of reaction (1):

$$MeOH_2^+ \xrightarrow{k'} Me^+ + H_2O_{(g)} \tag{7}$$

Since reaction (7) is an elementary process with rate k', according to the molecularity of this reaction, the rate expression is the following:

$$-\frac{d[MeOH_2^+]}{dt} = k' \cdot [MeOH_2^+] \tag{8}$$

After separation of variables, Eq. (8) can be rewritten as follows:

$$d\ln[MeOH_2^+] = -k' \cdot dt \tag{9}$$

that gives by integration:

$$\ln\frac{[MeOH_2^+]}{[MeOH_2^+]_0} = \ln\frac{I}{I_0} = -k' \cdot t \tag{10}$$

Eq. (10) allows to calculate the specific rate (k') of the dehydration reaction by using high-frequency AC current intensity measurements (see Figure 6). It was found a value for k' of (204 ± 2.6·10−4)min$^{-1}$ for dehydration in dry air.

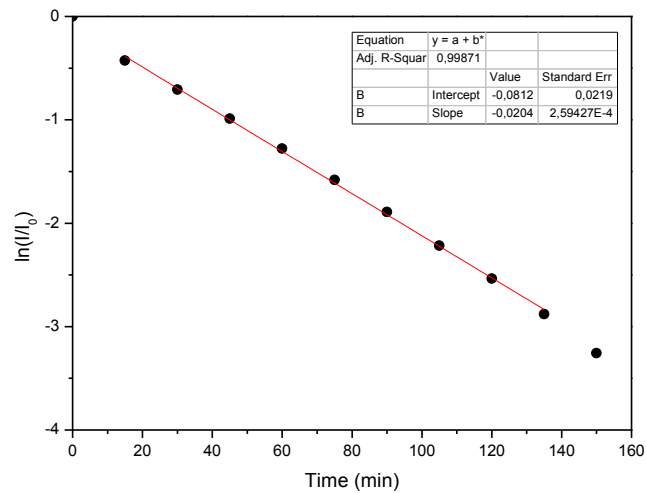

**Fig. 6** - Temporal evolution of ln(I/I$_0$) for the clinoptilolite spontaneous dehydration in dry air.

When clinoptilolite desorbs water in air, simultaneous hydration is possible inside the channels. The water concentration can be considered as a constant (environmental humidity) and for a fully hydrated sample at dehydration beginning it results: $[MeOH_2^+] \sim [Me^+]_0$, then the reaction kinetics can be approximately described by a first-order:

$$\frac{d[MeOH_2^+]}{dt} = k \cdot [H_2O] \cdot ([Me^+]_0 - [MeOH_2^+]) - k' \cdot [MeOH_2^+] \sim - k' \cdot [MeOH_2^+] \qquad (11)$$

As visible in Figure 7, at beginning of dehydration in air the kinetic constant has a value slightly smaller than the kinetic constant for dehydration in dry air (ca. $(70 \pm 2.6) \cdot 10^{-4}$min$^{-1}$).

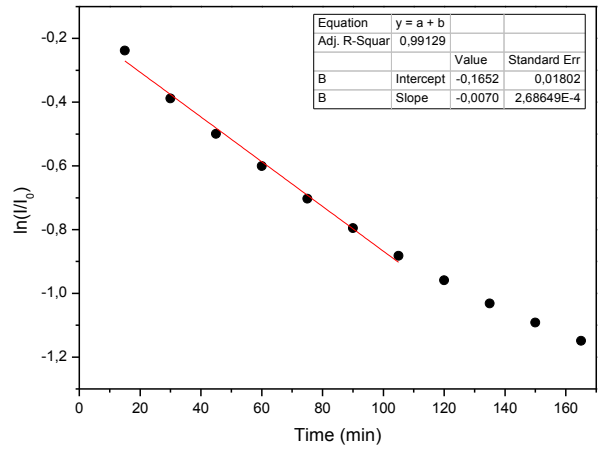

**Fig. 7** - Temporal evolution of $\ln(I/I_0)$ for the clinoptilolite spontaneous dehydration in air.

The equilibrium constant, $K_{eq}$, of the water hydration reaction ($Me^+ + H_2O = MeOH_2^+$) at room temperature is given by the ratio of the direct and inverse kinetic constants $K_{eq}=k/k'$ and it is 0.152. Therefore, the free energy variation in standard conditions ($\Delta G_0 = -R \cdot T \cdot \ln(k_{eq})$, where T is the absolute temperature and R the universal gas constant) of the hydration reaction is +4.66kJ/mol, that indicates a non spontaneous process in dry air (dehydration is a spontaneous process, $K_{dehydration}=1/K_{eq}=6.58$).

It is possible to formulate a model for the water adsorption/desorption process which is in accordance with the achieved kinetic results. In this model, the water molecules are first captured by the electric field of cations that is able to act at a distance longer than the dipole-dipole interaction possible between water and framework negative charges (short range interactions). The water molecules associated with cations may stay in two different configurations, each one with a characteristic energy content. Precisely, water molecules can stay at framework-cation interface and in this case the framework-cation interaction ($E_{f-c}$) is weak, but there are: hydrogen bonds between the framework negative charges and the water molecules ($E_{f-w}$), and water dipole-cation interactions ($E_{w-c}$). On the other hand, water molecules can be not interposed between framework and cation and in this case the cation-framework electrostatic interaction is strong, but there are not hydrogen bonds between water and framework negative charges, in addition to the water dipole-cation interactions. Probably the cationic site can flip between these two configurations having a different energy content and, depending on the environmental humidity (presence or absence of gaseous water molecules), one configuration could be energetically favored compared to the other. In fact, in presence of humidity (e.g., 75% moisture) the configuration with interposed water molecules may result energetically favored because further gaseous molecules can associate with cation by dipole-cation interactions, with lowering of the system energy content ($E_{f-w}+(n+1) \cdot E_{c-w} > E_{f-c}+E_{c-w}$). In these conditions, water adsorption is favored because it lowers the system total energy content. However, in absence of humidity (i.e., in air dried by silica gel) the energetically favored configuration of cationic sites should become that with non interposed water molecule, since in this case a stronger electrostatic interaction between cation and framework negative charges is possible ($E_{f-c}+E_{c-w} > E_{f-w}+E_{c-w}$). Differently from the 'interposed water configuration', the last site configuration allows desorption of the water molecules from the cationic site.

In this study, water has been selected as probe molecule to investigate the mechanism of adsorption because it is quite easy to generate atmospheres with a constant and exactly known water content, however a similar mechanism should be active also in the case of other dipolar molecules having comparable size and dipole moment (e.g., $H_2S$, $CH_3OH$, $H-CHO$, etc.).

When the dynamic equilibrium between water in the vapor phase and water in the framework cavity system is perturbed (under isothermal conditions), a very prompt change in the conduction properties can be observed. Such a perturbation can be simply produced, for example, by breathing close to the clinoptilolite-K sample. The prompt variation of electrical conductivity is a consequence of the direct cation-water interaction. As visible in Figure 8, the clinoptilolite-K electrical conductivity (observed as an increase in the AC current intensity) grows during each inhalation stage, but the material conductivity rapidly recovers the starting value during the exhalation step. Therefore, this material can be used to fabricate fast-responding sensors to count the human breathing events or to measure the human breathing rate [19]. Such a property can be advantageously exploited for technological applications in medicine (low-cost personal spirometry, general anesthesia monitoring, etc.) and sport fields [20]. Obviously, the above described water absorption/desorption kinetic controls, active for the clinoptilolite-K cations, takes place also during the operation of this type of humidity sensor, in fact, as visible in Figure 8, each breathing detection step consists of successive water adsorption and desorption reactions.

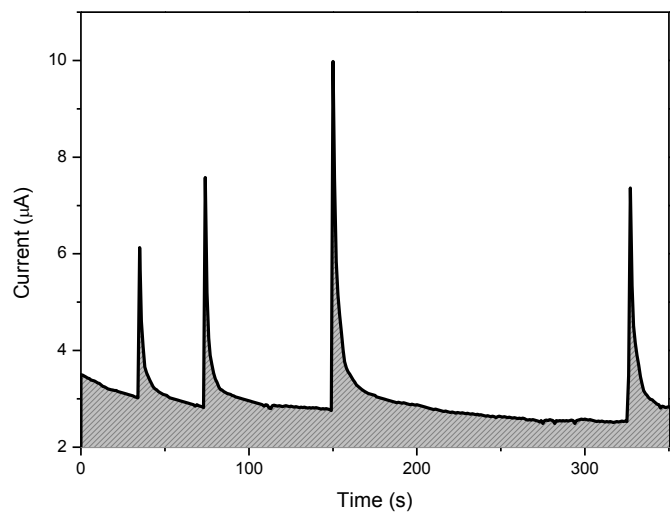

**Fig. 8 -** Sequence of four couples of water adsorption and desorption processes taking place in the clinoptilolite-K sample exposed to human breathing.

The maximum current intensity achieved during the human breathing monitoring is related to the deepness of breathing event.

## 4. Conclusions

An original approach for the selective investigation of water adsorption on cationic sites has been developed. This approach is based on monitoring water adsorption by an electrical technique (increase of current intensity in a sample biased by a constant high-frequency AC voltage), which gives information on the behavior of the only cationic sites. To keep out all non-cationic sites (external and framework Si-OH) from the adsorption process, the clinoptilolite sample has been first exposed to a hydration/dehydration cycle which, for the faster dehydration of cationic sites, allowed saturation by water of all non-cationic sites (silanols, Si-OH). According to the kinetic investigation made by this approach, cations are involved in the clinoptilolite hydration/dehydration processes and water adsorption/desorption takes place by the addition of single water molecules to each cation present in the channels. The dehydration reaction has been found much faster than the hydration reaction and water desorption from cations is a spontaneous process both in air and dry atmosphere for the investigated clinoptilolite sample. The increase of cation mobility following to its association with the water molecule could be explained by assuming that the water molecules attracted by the cation electric field settle down between cation and one of the negatively charged oxygen atoms present in the cationic site. The possibility to use this material to fabricate fast-responding humidity sensors useful for example for the human breathing monitoring has been also described.

**References**

1. I. M. Kalogerasand and A. Vassilikou-Dova, "Electrical properties of zeolitic catalysts," *Defect and Diffusion Forum*, vol. 164, pp.1-36, 1998.

2. T. Armbruster, "Dehydration mechanism of clinoptilolite and heulandite: single-crystal X-ray study of Na-poor, Ca-, K-, Mg-rich clinoptilolite at 100K," *American Meneralogist,* vol. 78, pp. 260-264, 1993.

3. S. Reib, G. Hagen, and R. Moos, "Zeolite-based impedimetric gas sensor device in low-cost technology for hydrocarbon gas detection," *Sensors,* vol. 8, pp. 7904-7916, 2008.

4. M. P. Pina, R. Mallada, M. Arruebo, M. Urbiztondo, N. Navascues, O. de la Iglesia, and J. Santamaria, "Zeolite films and membranes. emerging applications," *Microporous and Mesoporous Materials*, vol. 144, pp. 19-27, 2011.

5. A. A. Christy, "The nature of silanol groups on the surfaces of silica, modified silica and some Silica based materials," *Advanced Materials Research,* vol. 998-999, pp. 3-10, 2014.

6. G. Carotenuto, "How 'hydrophilic sites' work in the water adsorption/desorption by Natural Clinoptilolite," *European Journal of Engineering Research and Science,* vol. 4(3), pp. 183-189, 2019.

7. H. Lin, Q. Liu, Y. Dong, Y. Chen, H. Huo, and S. Liu, "Study on channel features and mechanism of clinoptilolite modifies by $LaCl_3$," *Journal of Materials Science Research*, vol. 2(4), pp. 37-44, 2013.

8. C. Cobzaru, A. Marinoiu, and C. Cernatescu, "Sorption of vitamin C on acid modified clinoptilolite," *Rev. Roum. Chim.*, vol. 60(2-3), pp. 241-247, 2015.

9. "Zeolite characterization and catalysis," A. W. Chester, E. G. Derouane Eds., Springer, Heidelberg (Germany), 2009.

10. N. Mansouri, N. Rikhtegar, H. A. Panahi, F. Atabi, and B. K. Shahraki, "Porosity, characterization and structural properties of natural zeolite - Clinoptilolite - as a sorbent," *Environmental Protection Engineering,* vol. 39(1), pp. 139-152, 2013.

11. W. Rodòn, D. Freire, Z. de Benzo, A. B. Sifontes, Y. Gonzalez, M. Valero, and J. L. Brito, "Application of a 3A zeolite prepared from venezuelan kaolin for removal of Pb(II) from wastewater and its determination by flame atomic absorption spectrometry," *American Journal of Analytical Chemistry*, vol. 4, pp. 584-593, 2013.

12. M. Stylianou, V. Inglezakis, A. Agapiou, G. Itskos, A. Jetybayeva, and M. Loizidou, "A comparative study on phyllosilicate and tectosilicate mineral structural properties," *Desalination and Water Treatment*, vol. 112, pp. 119-146, 2018.

13. D. Nibou, S. Amokrane, H. Mekatel, and N. Lebaili, "Elaboration and characterization of solid materials of types zeolite NaA and faujasite NaY exchanged by zinc metallic ions $Zn^{2+}$," *Physics Procedia,* vol. 2, pp. 1433-1440, 2009.

14. P. J. Launer and U. B. Arkles, "Infrared analysis of organosilicon compounds: spectra-structure correlations," in Silicon Compounds: Silanes and Silicones, Gelest, Inc., Morrisville, PA., 2013, pp. 175-178.

15. X. Xu, J. Wang, and Y. Long, "Zeolite-based materials for gas sensors," *Sensors,* vol. 6, pp. 1751-1764, 2006.

16. Y. Zheng, X. Li, and P. K. Dutta, "Exploitation of unique properties of zeolites in the development of gas sensors," *Sensors,* vol. 12, pp. 5170-5194, 2012.

17. I. Yimlamai, S. Niamlang, P. Chanthaanont, R. Kunanuraksapong, S. Changkhamchom, and A. Sirivat, "Electrical conductivity response and sensitivity of ZSM-5, Y, and mordenite zeolites towards ethanol vapor," *Ionics,* vol. 17, pp. 607-615, 2011.

18. O. Schaf, H. Ghobarkar, F. Adolf, and P. Knauth, "Influence of ions and molecules on single crystal zeolite conductivity under in situ conditions," *Solid State Ionics,* vol. 143, pp. 433-444, 2001.

19. U. Mogera, A. A. Sagade, S. J. George, and G. U. Kulkarni, "Ultrafast response sensor using supramolecular nanofibre and its application in monitoring breath humidity and flow," *Scientific Reports,* vol. 4:4103, pp. 1-9, 2014.

20. M. Niesters, R. Mahajan, E. Olofsen, M. Boom, S. Garcia del Valle, L. Aarts, and A. Dahan, "Validation of a novel respiratory rate monitor based on exhaled humidity," *Brithis Journal of Anaesthesia,* vol. 109(6), pp. 981-9, 2012.